\documentclass[review]{elsarticle}

\let\today\relax
\makeatletter
\def\ps@pprintTitle{%
    \let\@oddhead\@empty
    \let\@evenhead\@empty
    \def\@oddfoot{\footnotesize\itshape
         {} \hfill\today}%
    \let\@evenfoot\@oddfoot
    }
\makeatother

\usepackage{amssymb}
\usepackage{amsmath,amsfonts}
\usepackage{setspace}
\usepackage{tocloft}
\usepackage{xcolor}
\usepackage{array}
\usepackage{multirow}
\usepackage{algorithm}
\usepackage{algorithmic}
\usepackage{enumitem}
\usepackage{mathtools}
\usepackage{hyperref}
\usepackage{changepage}
\usepackage{amssymb}
\usepackage{graphicx}
\usepackage{textcomp}
\usepackage{booktabs}
\usepackage{url}
\usepackage{color}
\usepackage{caption}
\usepackage[figuresright]{rotating}

\newcommand{\tabincell}[2]{\begin{tabular}{@{}#1@{}}#2\end{tabular}}

\journal{}

\begin{document}

\begin{frontmatter}

\title{Medical Knowledge-Guided Deep Learning for Imbalanced Medical Image Classification}

\author[a,b]{Long Gao}
\author[c]{Chang Liu}
\author[a]{Dooman Arefan}
\author[d]{Ashok Panigrahy}
\author[a,e]{Margarita L. Zuley}
\author[a,c,f,g]{Shandong Wu\footnote{Corresponding author: Shandong Wu, wus3@upmc.edu}}

\address[a]{Department of Radiology, School of Medicine, University of Pittsburgh, 4200 Fifth Ave, Pittsburgh PA, 15260, USA}
\address[b]{College of Computer, National University of Defense Technology, 410073, Changsha}
\address[c]{Department of Bioengineering, Swanson School of Engineering, University of Pittsburgh, 4200 Fifth Ave, Pittsburgh PA, 15260, USA}
\address[d]{Department of Radiology, University of Pittsburgh, University of Pittsburgh Medical Center Children's Hospital of Pittsburgh, 4401 Penn Ave, Pittsburgh, PA, 15224, USA}
\address[e]{Magee-Womens Hospital of University of Pittsburgh Medical Center, 300 Halket St, Pittsburgh, PA, 15213, USA}
\address[f]{Department of Biomedical Informatics, University of Pittsburgh, 4200 Fifth Ave, Pittsburgh PA, 15260, USA}
\address[g]{Intelligent Systems Program, University of Pittsburgh, 4200 Fifth Ave, Pittsburgh PA, 15260, USA}

\begin{abstract}
Deep learning models have gained remarkable performance on a variety of image classification tasks. However, many models suffer from limited performance in clinical or medical settings when data are imbalanced. To address this challenge, we propose a medical-knowledge-guided one-class classification approach that leverages domain-specific knowledge of classification tasks to boost the model's performance. The rationale behind our approach is that some existing prior medical knowledge can be incorporated into data-driven deep learning to facilitate model learning. We design a deep learning-based one-class classification pipeline for imbalanced image classification, and demonstrate in three use cases how we take advantage of medical knowledge of each specific classification task by generating additional “middle” classes to achieve higher classification performances. We evaluate our approach on three different clinical image classification tasks (a total of 8459 images) and show superior model performance when compared to six state-of-the-art methods.  
\end{abstract}


\end{frontmatter}

\section{Introduction}
\label{Intro}

Medical imaging classification tasks have achieved remarkable performance with the development of deep learning models. Usually, the training of most deep learning models requires a large number of samples from balanced classes. However, in many medical or clinical applications, samples belonging to certain classes are hard to collect or rare due to low prevalence, resulting in data being imbalanced, which poses a great challenge for building machine learning models \cite{litjens2017survey}.

To address the data imbalance problem, One-Class Classification (OCC) is proposed to learn a model from the class that has a majority of samples, while other classes that have minority of samples are considered as the anomaly to be detected or separated \cite{johnson2019survey,ruff18deepsvdd}. This is also similar to the problem of anomaly detection   \cite{ruff18deepsvdd,schlegl2017unsupervised}. For the OCC, the key point is to build a model that performs well on the majority class samples, while under-performs on other class samples.  Classical methods, such as One-Class Support Vector Machine (OCSVM) \cite{scholkopf2000support} and Support Vector Data Description (SVDD) focus on mapping features \cite{tax2004support}, they are hard to handle large-scale datasets \cite{ruff18deepsvdd}. 
On the other hand, deep learning-based methods, such as autoencoder \cite{baur2018deep,chen2018unsupervised} and generative adversarial network \cite{Goodfellow2014Generative}, focus on modeling the distribution of the majority class samples. However, these methods are purely data-driven, leading to highly varying effects in different medical image classification tasks.
For deep learning-based medical image analysis, there is emerging but limited work that attempts to integrate prior knowledge to improve machine learning  \cite{milletari2017integrating}. Some studies \cite{konishi2000statistical,madabhushi2003combining} introduce color and texture knowledge to obtain regions of interest for image segmentation. To date, there is little to none previous work that incorporates specific prior medical knowledge into OCC for classifying imbalanced medical images.

In this paper, we propose a novel end-to-end method, namely, $M$edical $K$nowledge-guided $D$eep $L$earning (MKDL), for classifying imbalanced medical images by one-class classification. The principle behind MKDL lies in the fact that many medical image classification tasks are partly or largely based on certain pre-known image features, such as image contrast and brightness. These pre-known features can be computationally represented and then leveraged as prior knowledge to facilitate more effective model learning and therefore boost the classification performance. We formulate this idea in the self-supervised learning  \cite{kolesnikov2019revisiting} framework, where we construct a balanced classification task with newly generated "middle" classes through transforming the original samples of the “majority” class into additional image sample counterparts. After the classifier is trained on a mixture of the original and transformed image samples, it learns to distinguish features on samples from the “majority” class while under-performs on samples from the “minority” class, making the two classes of the data distinguishable.

The contributions of this work are summarized as follows: 

$\cdot$ We propose a new deep learning approach that can incorporate domain-specific prior medical  knowledge to improve the classification tasks on imbalanced imaging data.

$\cdot$ We propose a simple but effective transformation method to utilize task/data-specific knowledge into deep learning process.

$\cdot$ We demonstrate our approach on three different use cases where specific knowledge is reflected by the proposed transformations and outperforming classification performance are shown in comparison to six related methods.

\section{Related Work}
\label{related}

Our work focuses on integrating medical knowledge into deep learning models in OCC to deal with imbalanced medical image data classification. So here we mainly review some of the major related work on imbalanced data classification and prior-knowledge-guided machine learning.

$\textbf{Imbalanced Data Classification }$  Imbalanced data classification can be dealt with two schemes: one-class classification or binary classification with special strategies \cite{johnson2019survey}. The first scheme is also known as anomaly detection \cite{aldweesh2020deep,lv2020layer}. In this scheme, a model is trained with samples from the majority class, while samples from the minority class are regarded as abnormalities to be detected by the trained model.   The classical methods usually rely on mapping features into new feature space \cite{scholkopf2000support} by functions or fitting the distribution of majority samples \cite{zong2018dagmm}. For example, OCSVM and SVDD employ kernel functions to map input features into new hyperplane/hypersphere. These methods are limited for large scale data because of the curse of dimensionality \cite{ruff18deepsvdd}.  With the development of deep learning, convolutional neural networks (CNN) also have been widely applied in anomaly detection. A straightforward method is to train an autoencoder with normal samples, and classify testing samples by analyzing the reconstruction loss \cite{baur2018deep,chen2018unsupervised}. Some methods also integrate adversarial learning \cite{Goodfellow2014Generative} and other strategies  \cite{zhou2006extraction,zong2018dagmm} to enhance the performance \cite{wei2018anomaly,schlegl2017unsupervised,tang2019chest}. For example, AnoGAN \cite{schlegl2017unsupervised} learns the distribution of normal samples by training a GAN. DAOL optimizes autoencoder by a GAN model to fit the normal chest X ray samples. Similar structure is also adopted by ADOCC \cite{wei2018anomaly} for CT and PET images. These reconstruction-loss-based methods are relatively difficult to train, and the computational cost increases when dealing with large scale datasets. 

When using binary classifiers, the key is to balance the influence of different class samples in the binary classification models \cite{bulo2017loss,viola2001rapid}. A straightforward method is to balance the sample numbers by over-sampling or under-sampling \cite{de2019boosting,viola2001rapid}. Another method is to generate more minority class samples by sampling from the borderline area  \cite{han2005borderline,nguyen2011borderline}. Some previous studies also use cost-sensitive methods \cite{zadrozny2003cost} such as focal loss \cite{lin2017focal} and CosNet \cite{bertoni2011cosnet} to manipulate weights on the samples based on the cost metrics.

$\textbf{Prior-knowledge-guided learning}$ Integrating prior knowledge is a broad concept which has been intensively studied in different contexts. Some previous work using prior knowledge focuses on image segmentation, detection, and classification. For instance, PCA-CNN \cite{milletari2017integrating} proposes a PCA-based layer to enforce the model to focus on specific regions for segmentation tasks.   \cite{zhang2019knowledge} proposes a method to integrate prior knowledge to minimize the distribution distance of different domains for segmentation. \cite{chai2018glaucoma} combines the information about important regions to improve the performance of glaucoma diagnosis. The work of \cite{guan2009lung} introduces gene features into lung cancer binary classification task to improve the performance. \cite{liu2019automated} proposes a CNN model to integrate the nodule size and shape distribution as a prior knowledge for thyroid nodule detection in ultrasound images. As far as we know, little work attempts to use prior medical knowledge in the context of the OCC task on medical images. A previous work that is somewhat related to using knowledge for OCC is \cite{wang2018hyperparameter}. This paper generates negative samples by shifting edge points along the negative direction of estimated data density gradient, thus selects optimal hyper-parameters for OCSVM.

\section{Proposed methods}

\subsection{Task formulation}

We focus on the classification problems on imbalanced data by the One-Class Classification approach. In OCC, the class with the majority samples is considered as the normal class, while the samples in the minority class are considered abnormalities or outliers. It can be formulated as follows: for a set of samples $\mathcal{X}$ belonging to the majority class, we aim to learn a scoring function $F(\mathcal{X})$: $\mathcal{X}$ $\rightarrow$ $R$. A higher value of $F(X)$ indicates that the given sample $X$ has a higher likelihood of belonging to the majority class. 

\begin{figure*}[ht!]
\centering
\includegraphics[scale=0.31]{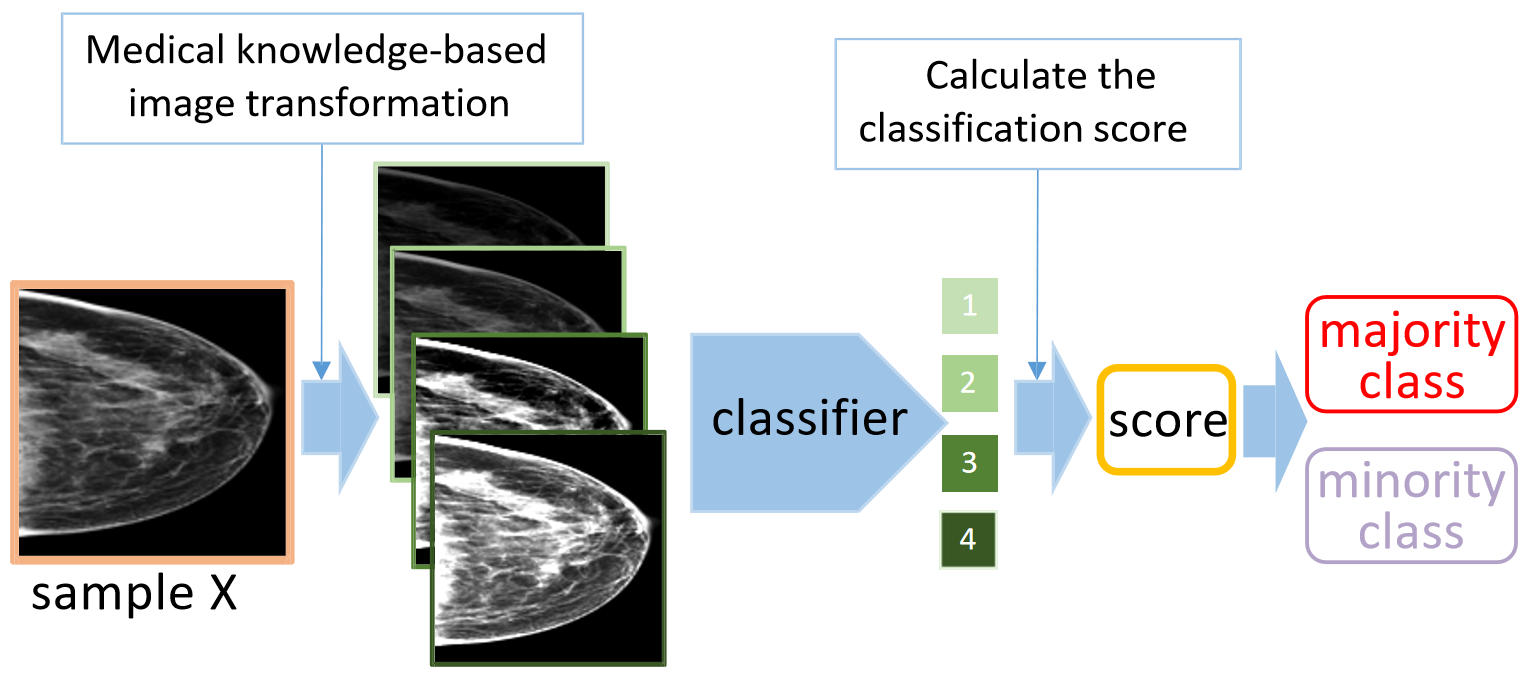}
\caption{\label{fig-structure} The pipeline of method. First, a given sample $X$ is transformed to generate n (here n=4 as an example) images based on medical knowledge. Then a classifier is trained to classify the transformed images into the n classes . Finally, we calculate the classification score ($s$) that represents the probability of sample $X$ belonging to the majority class.}
\end{figure*}

\begin{figure*}
\centering
\includegraphics[trim=0 260 0 80, scale=0.78]{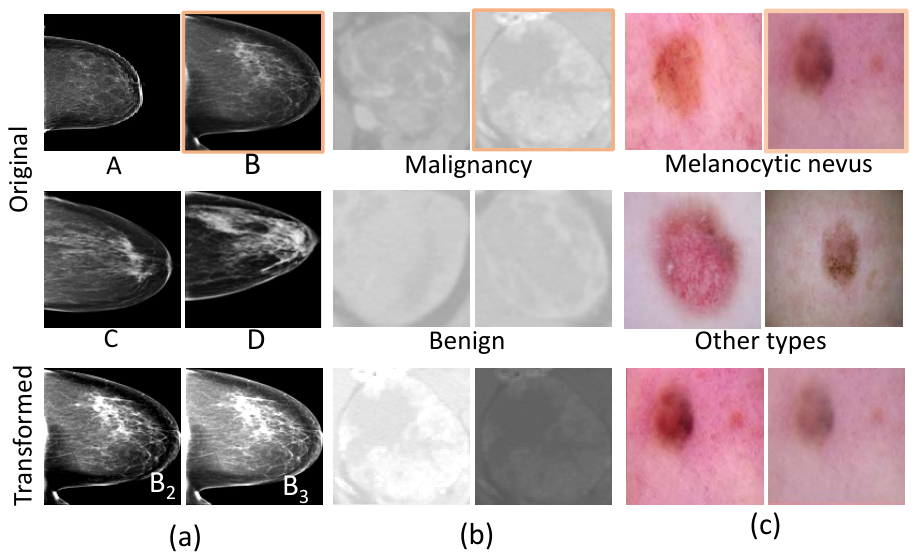}
\caption{\label{fig-demo-data} Demonstration images of the different classes (the first two rows) and transformed images (the third row, generated from the right-top sample marked with the yellow box) in the three use cases: (a) Breast density; (b) Kidney tumor; (c) Skin cancer.}
\end{figure*}

\subsection{Rationale}

With OCC, our goal is to construct a task, where the model can learn effectively on samples belonging to the majority class, while exhibiting a low recognition performance on samples belonging to the minority class. To this end, inspired by self-supervised learning \cite{kolesnikov2019revisiting}, we construct a classification task by training the model to classify a set of artificially transformed images into corresponding classes which represent the transformations. In this process, selecting appropriate image transformations is crucial. A series of transformations (e.g., rotation, shift \cite{golan2018DeepAD,kolesnikov2019revisiting}, etc) have been studied on natural images. However, different from natural images where the classes are usually relevant to different objects/scenes, the classification of medical/clinical images are often related to different image properties such as the brightness and contrast features. Based on this observation, a novel idea is to implement image transformations through manipulating the classification-related contrast and brightness features, guided by pre-known medical knowledge that is specific to an image classification task or dataset. These transformations generate additional samples (we denote them as of newly constructed “middle” classes, since these samples have a different distribution from the original majority class). Then, based on the original and the “middle” classes, we build a classifier to effectively learn the intrinsic image features of the majority class.


\subsection{Proposed pipeline}
\label{sec-pipeline}


The pipeline with proposed methods ( Figure \ref{fig-structure} )includes three steps as outlined in the following paragraphs.

First, generate n image counterparts from an original sample $X$ according to specific prior knowledge. See details in Section \ref{sec-explaining-intuition} for demonstration on three use cases.
details of using prior medical knowledge to generate these images are demonstrated in Section \ref{sec-explaining-intuition} on three use cases. 
Specifically, the generation of images is defined as: $\mathcal{T}=\{ \mathcal{T}_1, \cdots, \mathcal{T}_n\}$ with the corresponding class labels \{1, $\cdots$, n\}. $\mathcal{T}_i$=$\mathcal{T}_i(X)$ represents the $i$th transformation operation applied to an original sample $X$.  
A variety of operations may be used to implement the transformations. Considering that in many cases different classes of medical images are usually more relevant to the image contrast, brightness, and/or color features, in this work we use a simple linear transformation operation, namely, Linear Magnification (LM), as defined in Formula \ref{equ-transformation1}, to utilize  potential medical knowledge for classification.

\begin{equation}
\label{equ-transformation1}
\mathcal{T}_i = c \cdot \mathcal{T}_1 + b; \ \ \ \mathcal{T}_1=X.   
\end{equation}
where $\mathcal{T}_1$ is the original sample $X$; $c$ and $b$ denote the two coefficients adjusting the contrast and brightness of the sample, respectively. 

It is important to point out that even though the operation of the proposed transformation is similar to data augmentation, the aim of the transformation operation in our approach is fundamentally different from data augmentation \cite{shorten2019survey}. Data augmentation is to modify class-invariant features and preserve class-relevant features to keep the generated images in the same class as the original images. In contrast, the aim of the transformation operations in our approach is to modify the class-relevant features so that the generated images will have a different distribution from the original class. 

Second, train a classifier with the image-label pairs ($\mathcal{T}_i$, $i$) generated from the majority class samples through the first step.  Here we employ the Wide Residual Network (WRN)   \cite{zagoruyko2016wide} as the classifier, and use cross-entropy loss as the loss function.

Third, calculate the classification score for a given sample. The classifier will output a 1$\times$n vector for each input image. For the n images generated from $X$, we can obtain a matrix P$_{n\times n}$. P($i$, $j$) represents the probability of image $\mathcal{T}_i$ ($i\in$\{1, \dots, n\} belonging to class $j$ ($j\in$\{1, \dots, n\}). The classification score $s$ is calculated by the summation of the diagonal elements of matrix P (see Formula \ref{equ-score}). 
Since the classification model is trained on the generated images from the samples belonging to the “majority” class, a given test sample from the samples of the majority class will be well-recognized by the classifier. Hence P will be more like a unit diagonal matrix in this situation. On the other hand, when a test sample coming from the “minority” class is tested, the classifier will under-perform. Hence P will be less like a unit diagonal matrix. Based on this difference, a classification score $s$ is calculated as the summation of the diagonal elements of matrix P by Formula \ref{equ-score}; then by assessing the value of score $s$, a testing sample can be classified into either the majority or the minority class.

\begin{equation}
\label{equ-score}
s = \sum_{i=1}^{i=n} P(i,i)
\end{equation}

\subsection{Incorporating prior medical knowledge: three use cases}
\label{sec-explaining-intuition}

The transformation operations are a key to generate transformed images in order to build an effective classification model. Different transformation operations can be used or customized according to the characteristics of specific tasks/datasets. In medical fields, while there is more generic knowledge that can be useful across different fields, knowledge often is specific to a concrete task/dataset and therefore, it is vital to represent and implement specific knowledge for a specific use case. In this section, we demonstrate how prior medical knowledge is utilized to generate the additional samples of middle classes on three different use cases for image classification tasks. For each task, we specify the knowledge that is related to contrast and brightness of images, and then implement the knowledge through the proposed transformations.

$\textbf{Mammographic breast density category classification:}$ 
This task focuses on distinguishing the four Breast Imaging Reporting and Data System (BI-RADS)-based breast density categories (see Figure \ref{fig-demo-data}(a)) based on digital mammograms: A) almost entire fatty, B) scattered areas of fibroglandular density, C) heterogeneously dense, D) extremely dense. Category A and B are considered a lower risk of developing breast cancer, while C and D indicate a higher risk of developing breast cancer. Statistically, 10\% patients in breast cancer screening fall into category A and D, respectively, and 40\% patients fall into B and C, respectively. 
Thus, it forms a typical four-class classification problem with an imbalanced number of samples across the classes.

Medical  knowledge: The classification of BI-RADS density categories is generally based on the amount of dense/fibroglanduar tissue in the breast. Images belonging to Category C and D have more dense pixels/regions, resulting in a higher level of contrast and brightness. Therefore, this knowledge can be reflected by simply magnifying the pixel intensity values via implementing the LM transformation operation. An example is shown in the third row of Figure \ref{fig-demo-data}(a), where we generate two middle classes  B$_2$ and  B$_3$ by multiplying a constant on the pixel intensity of the image B: B$_2$ =1.2$\times$B, B$_3$=1.4$\times$B.  By training with image-label pairs: (B, 1), (B$_2$, 2) and (B$_3$, 3), the classifier can effectively learn the features related to the density category classification.

$\textbf{Kidney tumor diagnosis:}$ This task focuses on distinguishing the benign lesions (including benign and cyst) from malignant lesions (renal cell carcinoma) based on CT images. Because of the low prevalence of kidney carcinoma (i.e., malignancy is less often is general population) or benign (for example, in certain special circumstances such as in cancer specialty hospitals, there are many more carcinoma cases than benign), the classification of the two classes can be imbalanced.

Medical  knowledge: As shown in the first two rows of Figure~\ref{fig-demo-data}(b), the malignancy cases have higher contrast and more amounts of bright regions because they have more artery blood vessels revealed by the contrast agent.
Thus, if a classifier is trained on images generated from malignancy cases (an example is shown in the third row of Figure \ref{fig-demo-data}(b)), it will have low accuracy when distinguishing images generated from benign cases that have less amounts of contrast and brightness.

$\textbf{Melanocytic nevus diagnosis in skin cancer:}$ 
This task focuses on distinguishing different skin cancer types based on dermatoscopic images. The classification is often imbalanced because each type of skin cancer has a different prevalence rate. Melanocytic nevus (MN) is the most common skin cancer among other types of skin cancers such as melanoma, basal cell carcinoma, actinic keratosis, benign keratosis, dermatofibroma, and vascular lesion \cite{tschandl2018ham10000}. Here we aim to classify images as MN or other skin cancers.

Medical  knowledge: Color is an important distinguishing factor for the clinical diagnosis of skin cancer, especially for MN \cite{scope2008ugly}. As shown in the first two rows of Figure \ref{fig-demo-data}(c), MN lesions are usually manifested brown in images. Thus, by adjusting contrast and brightness, we can generate new images to train a classifier to learn the color characteristics associated with MN diagnosis.



In summary, these three clinical image classification tasks are usually challenging due to imbalanced data. Using contrast and brightness transformation operations guided by specific medical knowledge, we can generate additional images that have different distributions from the given class to facilitate self-supervised learning.

\section{Experiments}
\label{sec:Experiments}

In this section, we evaluate MKDL from four aspects: 1) we show the classification performance of MKDL with a fixed transformation operation and compare MKDL to other related methods (Section \ref{sec-comp-one}); 2) we  evaluate the performance of MKDL under different transformation operations with a varying combination of transformation parameters (Section \ref{sec-comp-ft}); 3) we analyse the influence of training sample number to show that MKDL is robust to the small scale training set (Section \ref{sec-comp-num}).

\subsection{Datasets}
\label{sec-dataset}

\begin{table*}[htbp]
\centering
\caption{Datasets statistics.}
\label{tab:data}
\begin{tabular}{c| c| c c| c| c c }
\toprule[0.6pt]
\multirow{2}{*}{Dataset} & \multirow{2}{*}{{Dimension}}&\multicolumn{2}{c|}{{Total}} & \multirow{2}{*}{Training}&\multicolumn{2}{c}{{Testing}} \\
&   & Majority & Minority &&Majority & Minority   \\
\midrule
\multirow{2}{*}{Breast Density}& \multirow{2}{*}{{128$\times$128$\times$1}} &B: 2058&A,C,D: 1690& B: 1500&558&1690\\
&&  C: 2261 & A,B,D: 1487 & C: 1500 & 761&1478\\
\midrule
Kidney Tumor&64$\times$64$\times$1&147 &32&115&32&32\\
\midrule
Skin Cancer&128$\times$128$\times$3&2022&978&1500&522&978\\
\bottomrule[0.6pt]
\end{tabular}
\end{table*}

Corresponding to the three use cases discussed in Section \ref{sec-explaining-intuition}, our experiments include three different clinical imaging datasets with different modalities, as briefly summarized in Table \ref{tab:data}. All images within a single dataset are resized to the same dimension, and all pixel intensity values are scaled to reside in [0, 1]. For the Breast Density dataset, identifying Category B and C is the most demanded and challenging task for radiologists because these two categories are similar to each other but indicate different clinical actions (i.e., category B for regular surveillance but category C for enhanced screening). Thus, we design two clinically-demanded experiments by 1) taking category B as the majority class and all other categories as the "minority" class, 2) taking category C as the majority class and all other categories as the "minority" class. In the Kidney Tumor dataset, the minority class has much fewer samples than the majority class.  
The Skin Cancer dataset is  publicly available \cite{tschandl2018ham10000}, we randomly select 3000 images for our experiments due to considerations on computational limitation.

\subsection{Baseline methods and experimental settings}
\label{sec-baseline}
 
The MKDL is compared to six previous OCC models: OCSVM \cite{scholkopf2000support}, C-OCSVM, DSEBM \cite{zhai2016deep}, DAGMM \cite{zong2018dagmm}, D-SVDD \cite{ruff18deepsvdd}, and AnoGAN \cite{schlegl2017unsupervised}. The details are as follows:

$\textbf{OCSVM}$: This method focuses on mapping the sample into a new feature hyperplane by a kernel function. Here the Radial Basis Function (RBF) kernel is used. RBF needs the hyper-parameter $\nu$ (anomaly ratio), so we grid search $\nu\in\{0.01, \cdots, 0.09\}$ to select the best performing values based on the ground truth labels. Therefore, what we report here is the upper bound of OCSVM's performance. Each image is reshaped into a vector as the input feature. 

$\textbf{C-OCSVM}$ This is a combination of CAE and OCSVM. We first train a CAE to compress images, then use the bottleneck features to train OCSVM. Similar to previous one, we also use grid search to find the best performance.

$\textbf{DSEBM}$ Deep Structure Energy-based Model uses an energy-based model to enforce the encoder to map samples from the majority class into a low-energy feature space. 

$\textbf{DAGMM}$ Deep Autoencoder Gaussian Mixture Model uses a Gaussian Mixture Model to help the autoencoder to escape from local optima and fit the distribution of training samples better.

$\textbf{D-SVDD}$ Similar to SVDD, Deep SVDD uses an encoder network to map the samples from the majority class into a new hypersphere. Here we use the default hyper-parameter settings for this method. 

$\textbf{AnoGAN}$ This is a GAN-based method for medical images. After training a GAN with samples from the majority class, it finds a latent code from the distribution to generate the most similar image to a given testing image, and use the difference between generated and original images as the loss. The data augmentation is also used in AnoGAN because the GAN needs a large number of samples to train.

We repeat each algorithm three times and report the average values of the model performance.
Keras \cite{chollet2015keras} is employed on an NVIDIA TITAN GPU. Adam \cite{Kingma2014Adam} is adopted as the optimizer, with a learning rate of 0.0002 and a batch size of 128. These parameters are fixed across all the experiments.  Two metrics are used to evaluate the classification performance: i) Area Under Receiver Operating Characteristic Curve (AUC), and ii) Area Under the Precision-Recall curve for majority class (AUPR-maj) and minority class (AUPR-min).

\begin{sidewaystable}
\centering
\caption{The AUC$\pm$std of different methods. The highest AUCs are marked in bold.} 
\label{tab:result}
\begin{tabular}{ c  c | c| c |c| c |c |c| c| c}
\toprule[0.6pt] 
\multicolumn{2}{c|}{Dataset}&Evaluate&OCSVM&C-OCSVM&DAGMM&DSEBM&D-SVDD&AnoGAN&MKDL\\
\midrule 
\multirow{6}{*}{\tabincell{c}{Breast\\ Density}}
&\multirow{3}{*}{B }&\multirow{1}{*}{\shortstack{AUC}} 
&74.4$\pm$1.1&69.5$\pm$6.2&58.7$\pm$8.7&70.0$\pm$8.7&59.2$\pm$9.5&69.8$\pm$3.7&\textbf{78.4$\pm$0.9}\\
\cline{3-10}
&&\multirow{1}{*}{\shortstack{AUPR-maj}} 
&40.5$\pm$0.5&40.1$\pm$7.0&38.7$\pm$15.8&40.9$\pm$7.2&24.4$\pm$5.0 &33.1$\pm$2.9&\textbf{56.3$\pm$4.9}\\
\cline{3-10}
&&\multirow{1}{*}{\shortstack{AUPR-min}} 
&88.5$\pm$0.8&85.9$\pm$3.6&87.3$\pm$1.5&87.2$\pm$5.4&74.8$\pm$8.1 &82.9$\pm$0.8&\textbf{90.5$\pm$0.2}\\
\cline{2-10}

&\multirow{3}{*}{C\ }&\multirow{1}{*}{\shortstack{AUC}} 
&56.4$\pm$1.4&65.6$\pm$3.8&49.8$\pm$17.9&54.2$\pm$3.3&59.2$\pm$4.5&60.7$\pm$8.5&\textbf{78.3$\pm$5.3}\\
\cline{3-10}
&&\multirow{1}{*}{\shortstack{AUPR-maj}} 
&29.4$\pm$0.6&44.2$\pm$2.5&36.2$\pm$10.4&50.8$\pm$13.7&35.6$\pm$7.6 &39.2$\pm$5.2&\textbf{58.1$\pm$8.5}\\
\cline{3-10}
&&\multirow{1}{*}{\shortstack{AUPR-min}} 
&63.1$\pm$0.9&79.5$\pm$4.0&40.0$\pm$1.0&69.3$\pm$4.5&68.9$\pm$8.1 &72.1$\pm$4.7&\textbf{89.2$\pm$2.4}\\
\midrule 

\multirow{3}{*}{\tabincell{c}{Kidney\\Tumor}}
&&\multirow{1}{*}{\shortstack{AUC}} 
&64.2$\pm$0.6&64.4$\pm$3.0&56.1$\pm$5.3&54.2$\pm$0.3&58.2$\pm$7.1&54.4$\pm$0.1&\textbf{67.0$\pm$3.1}\\
\cline{3-10}
&&\multirow{1}{*}{\shortstack{AUPR-maj}} 
&\textbf{63.2$\pm$0.4}&63.2$\pm$2.1&50.6$\pm$16.8&50.1$\pm$0.2& 43.7$\pm$3.8&50.3$\pm$0.7&63.0$\pm$3.3   \\
\cline{3-10}
&&\multirow{1}{*}{\shortstack{AUPR-min}} 
&70.0$\pm$0.4&67.5$\pm$4.8&61.7$\pm$11.5&61.6$\pm$0.5&49.9$\pm$5.9&61.7$\pm$0.1&\textbf{71.3$\pm$2.8}\\
\midrule

\multirow{3}{*}{{\tabincell{c}{Skin\\Cancer}}}&\multirow{3}{*}{}&\multirow{1}{*}{\shortstack{AUC}}
&67.4$\pm$1.3&62.9$\pm$5.3&55.1$\pm$5.7&52.0$\pm$4.4&61.7$\pm$1.5 &55.2$\pm$5.3&\textbf{72.8$\pm$1.8}\\
\cline{3-10}
&&\multirow{1}{*}{\shortstack{AUPR-maj}} 
&75.3$\pm$1.7&70.0$\pm$4.6&72.9$\pm$5.6&52.2$\pm$3.9&50.8$\pm$0.7&30.4$\pm$2.8&\textbf{78.5$\pm$1.7}\\
\cline{3-10}
&&\multirow{1}{*}{\shortstack{AUPR-min}} 
&52.2$\pm$1.5&61.9$\pm$5.0&38.9$\pm$7.6&50.9$\pm$3.1&33.3$\pm$1.0&64.2$\pm$2.1&\textbf{67.9$\pm$1.9}\\
\bottomrule[0.6pt]
\end{tabular}
\end{sidewaystable}

\subsection{Comparison with previous related methods}
\label{sec-comp-one}

In this experiment, we use a fixed transformation operation for all three datasets to show the “general” performance of our approach on the three different tasks. Note that there are four classes in the Breast Density dataset, categories C and D having higher contrast than B, and categories A and B having lower contrast than C. Taking this into consideration, we set n=5 for this task and we keep n=5 for the other two tasks too for consistency. Specifically, according to step 1 in Section \ref{sec-pipeline}, we transform each original sample into five images by setting the contrast coefficient $c$=\{0.2, 0.6, 1, 1.4, 1.8\} and the brightness coefficient $b$=\{0, 0, 0, 0, 0\} in Formula \ref{equ-transformation1}, where c=1, b=0 represents the original sample.  As shown in Table \ref{tab:result}, the proposed method outperforms the previous work in most of the experiments.

\subsection{Comparison of different transformation operations}
\label{sec-comp-ft}

\begin{table}[htbp]
\caption{\label{tab-different-trans}The AUCs under different transformation operations. The highest values are marked in bold. Note that LM(5,0) is the result reported in Table~\ref{tab:result} (the MKDL column).}
\centering
\begin{tabular}{c | p{0.3cm}<{\centering}|p{1.3cm}<{\centering}| p{1.2cm}<{\centering} p{1.2cm}<{\centering} |p{1.2cm}<{\centering} p{1.3cm}<{\centering}| p{1.3cm}<{\centering} | p{1.3cm}<{\centering}}
\toprule[0.6pt]
\multicolumn{2}{c|}{Dataset}&LM${(5,0)}$&S${(4,0)}$&R${(4,0)}$&LM${(5,1)}$&LM${(5,2)}$&LM${(3,0)}$&LM${(7,0)}$\\
\midrule 
\multirow{2}{*}{{Breast Density}}
&B&78.44&76.48&67.02&78.50&\textbf{81.28}&57.50&65.50\\
&C&78.30&67.20&70.39&82.61&\textbf{86.13}&66.99&77.89\\
\midrule 

\multicolumn{2}{c|}{\multirow{1}*{Kidney Tumor}} 
& 66.97&63.57&50.29&\textbf{74.67}&66.70&68.42&64.55\\
\midrule 

\multicolumn{2}{c|}{\multirow{1}*{Skin Cancer}} 
&72.81&69.23&67.47&\textbf{77.37}&76.41&71.31&76.45\\
\bottomrule[0.6pt]
\end{tabular}
\end{table}

In this part, we conduct the experiments using a different set of parameters for the LM transformation specifically customized for each task/dataset. To further reveal the benefits of using the medical knowledge guided transformations, we also compare MKDL with two straightforward transformation operations: shift (denoted by S(4,0)) and rotation (denoted by R(4,0)). In addition, to investigate the influence of the different number of transformation operations, we also test the performance when using n=3 and n=7 transformations (denoted by LM$(3,0)$ and LM$(7,0)$). The details of these different transformation operations are summarized as follows: 

$\textbf{S}(4,0)$: n=4, shift \{0, $\frac{h}{3}$\} pixels in each x and y directions for an image of size h$\times$h.

$\textbf{R}(4,0)$ \ \ : n=4, rotate \{0, 90$^\circ$, 180$^\circ$, 270$^\circ$\}.

$\textbf{LM}(5,1):$ n=5, c=$\{$0.6, 0.8, 1, 1.2, 1.4$\}$, 
						b=$\{$0.2, -0.2, 0, 0.2, -0.2$\}$.

$\textbf{LM}(5,2):$ n=5, c=$\{$0.6, 0.8, 1, 1.2, 1.4$\}$, b=$\{$0.4, -0.4, 0, 0.4, -0.4$\}$.

$\textbf{LM}(3,0):$ n=3, c=$\{$0.8, 1, 1.2$\}$, b=$\{$-0.2, 0, 0.2$\}$.

$\textbf{LM}(7,0):$ n=7,  c=$\{$0.4, 0.6, 0.8, 1, 1.2, 1.4, 1.6$\}$, b=$\{$0.2, -0.2, 0, 0.2, -0.2$\}$.

As shown in Table \ref{tab-different-trans}, the shift (S${(4,0)}$) and rotation (R${(4,0)}$) transformation operations obtain relatively good performance in some tasks, validating the proposed pipeline is a powerful method for AD. However, they  have lower performances than the proposed transformation operations (LM) that leverage prior medical knowledge. Particularly, the model with the rotation transformation collapses on the Kidney Tumor dataset (AUC=50.29) because the lesions have no specific direction property, indicating that transformations without the guidance of related knowledge can lead to failures. In contrast, with the prior medical knowledge as we specified, the models exhibit superior performances in the three given datasets. Furthermore, compared to LM${(5,0)}$, the optimized transformation parameter values customized for each task/dataset (LM${(5,1)}$ and LM${(5,2)}$) lead to increased performance.

\subsection{Comparison with varying size of training samples}
\label{sec-comp-num}

\begin{table*}[htbp]
\caption{\label{tab-different-num}The AUCs under a varying size of training sample.}
\centering
\begin{tabular}{c|c|p{0.6cm}<{\centering}p{0.6cm}<{\centering}p{0.6cm}<{\centering}p{0.6cm}<{\centering}}
\toprule[0.6pt]
\multicolumn{2}{c|}{Dataset}&1500&500&250&100\\
\midrule
Breast Density&MKDL&81.28&80.63&80.23&76.83\\
\cline{2-6}
Category B&OCSVM&74.39&74.13&74.27&73.62\\
\midrule 
Breast Density&MKDL&86.13&74.20&67.58&66.56\\
\cline{2-6}
Category C&C-OCSVM&65.60&65.83&56.71&54.65\\
\midrule 
\multirow{2}{*}{{Skin Cancer}}
&MKDL&76.41&73.16&73.35&71.96\\
\cline{2-6}
&OCSVM&67.40&66.85&65.94&66.87\\
\bottomrule[0.6pt]
\end{tabular}
\end{table*}

In this experiment we examine the effects of the MKDL method when training the models with a varying size of training samples. We report the model performance when using 100, 250, 500 samples for training. We perform the experiment on breast density and skin cancer dataset and skip this experiment on the kidney tumor dataset because it is already a quite small dataset. We use LM(5,2) as the transformation operation in implementing MKDL. We compare MKDL to other methods that obtain the best performance in the previous experiment (see Table 2). 
As shown in Table~\ref{tab-different-num}, the performance of MKDL is more stable than other methods with respect to the change of training sample size, and MKDL consistently outperforms other methods. In particular, the AUCs when using the smallest training dataset (100 samples) for MKDL are still higher than the highest AUCs of OCSVM and C-OCSVM when they are trained using 1500 samples.

\section{Conclusion}
\label{Conclusion---new}
 
In this work, we aim to address classification of imbalanced medical image data and we propose a novel one-class classifier that leverages medical knowledge to guide the deep learning of image features. By transforming the class-relevant features of a given sample into different images, we construct a classification task where the model can effectively learn essential information about the majority class. Our method achieves superior performance compared to previous one-class methods and other transformations, indicating that integrating medical knowledge into deep learning models may be an effective approach to improve data-driven machine learning. 
In future work, we will further explore the generalizability of our methods on diverse datasets from different machines and sites. While the medical knowledge is generic, different datasets may demand adaptive implementation of our methods.

\section*{Acknowledgment}
This project was supported in part by National Institutes of Health (NIH)/ National Cancer Institute (NCI) grant 1R01CA218405, the grant 1R01EB032896 as part of the National Science Foundation (NSF)/NIH Smart Health and Biomedical Research in the Era of Artificial Intelligence and Advanced Data Science Program, the UPMC Hillman Cancer Center Developmental Pilot Program, and a Developmental Pilot Award of the Pittsburgh Center for AI Innovation in Medical Imaging and the associated Pitt Momentum Funds through a Scaling grant from the University of Pittsburgh (2020). We gratefully acknowledge the support of NVIDIA Corporation for the donation of the Titan X Pascal GPU for our research. Most of the work was conducted when the first author was a visiting student in the University of Pittsburgh, with no funding support provided to the first author. The content of this study is solely the responsibility of the authors and does not necessarily represent the official views of the NIH or the NSF.

\section*{Disclosure}
Dr. Shandong Wu is a scientific consultant and stockholder of COGNISTX, Inc. Dr. Shandong Wu has a research grant funded by Amazon. All other authors have no conflicts of interests to disclose. 

\bibliography{PKDL}

\end{document}